\documentclass[pra,aps,twocolumn,nopacs,superscriptaddress,nofootinbib]{revtex4-1}

\usepackage{amsmath}  \usepackage{amssymb}  \usepackage{amsfonts}  \usepackage{bm}  \usepackage{bbm}   \usepackage{bbold}   \usepackage{braket}  \usepackage{color}  \usepackage{comment}  \usepackage{dcolumn}  \usepackage{enumerate}  \usepackage{epsfig}  \usepackage{gensymb}  \usepackage{graphicx}  \usepackage{indentfirst}  \usepackage{lmodern}  \usepackage{mathrsfs}  \usepackage{mathtools}  \usepackage{psfrag}  \usepackage{pst-all}  \usepackage{soul}  \usepackage{xcolor}
\usepackage{float}
\usepackage[colorlinks,linkcolor=blue,citecolor=blue,urlcolor=blue,hyperindex,driverfallback=dvipdfm]{hyperref}  \usepackage[T1]{fontenc}

\def\ii{{\rm i}}  \def\ee{{\rm e}}
\def\me{m_{\rm e}}  
  
\def\Ab{{\bf A}}              \def\fb{{\bf f}}            \def\jb{{\bf j}}    \def\kb{{\bf k}}            \def\qb{{\bf q}}    \def\rb{{\bf r}}       
\def\xx{\hat{\bf x}}  \def\yy{\hat{\bf y}}  \def\zz{\hat{\bf z}}        \def\eh{\hat{\bf e}}    
\def\EF{{E_{\rm F}}}     
    \def\kp{k_{\rm p}}     
        
\def\fb{{\bf f}}       \def\op{\omega_{p}}   \def\og{\omega_{\gamma}} \def\kbg{\kb_\gamma}   \def\fb{{\bf f}}   
\def\ii{{i}}  \def\ee{{e}}

\begin{document}

\title{Nanophotonics for pair production}

\author{Valerio~Di~Giulio}
\affiliation{ICFO-Institut de Ciencies Fotoniques, The Barcelona Institute of Science and Technology, 08860 Castelldefels (Barcelona), Spain}
\author{F.~Javier~Garc\'{\i}a~de~Abajo}
\email{javier.garciadeabajo@nanophotonics.es}
\affiliation{ICFO-Institut de Ciencies Fotoniques, The Barcelona Institute of Science and Technology, 08860 Castelldefels (Barcelona), Spain}
\affiliation{ICREA-Instituci\'o Catalana de Recerca i Estudis Avan\c{c}ats, Passeig Llu\'{\i}s Companys 23, 08010 Barcelona, Spain}

\begin{abstract}
{\bf The transformation of electromagnetic energy into matter represents a fascinating prediction of relativistic quantum electrodynamics that is paradigmatically exemplified by the creation of electron-positron pairs out of light. However, this phenomenon has a very low probability, so positron sources rely instead on beta decay, which demands elaborate monochromatization and trapping schemes to achieve high-quality beams. Here, we propose to use intense, strongly confined optical near fields supported by a nanostructured material in combination with high-energy photons to create electron-positron pairs. Specifically, we show that the interaction between near-threshold $\gamma$-rays and polaritons yields higher pair-production cross sections, largely exceeding those associated with free-space photons. Our work opens an unexplored avenue toward generating tunable pulsed positrons at the intersection between particle physics and nanophotonics.}
\end{abstract}

\maketitle

\section{Introduction}

The creation of massive particles from electromagnetic energy emerged as a prominent focus of attention in 1934, when the materialization of an electron and its antiparticle --the positron-- was predicted to occur with nonvanishing probability by Breit and Wheeler (BW) from the scattering of two photons \cite{BW1934}, by Bethe and Heitler (BH) from the interaction of a photon and the Coulomb potential of a nucleus \cite{BH1934}, and by Landau and Lifshitz (LL) from the collision of two other massive particles \cite{LL1934}. A main difference between these processes relates to the real or virtual nature of the involved photons. While only real electromagnetic quanta lying inside the light cone (i.e., satisfying the light dispersion relation in vacuum, $k=\omega/c$) participate in the BW mechanism for pair production, the LL process is mediated by two virtual photons, and both real and virtual photons play a role in BH scattering. Eventually, pair production was achieved by colliding energetic electrons and real photons delivered by high-power lasers \cite{BFH97}, and more recently using only real photons generated from atomic collisions \cite{AAA21}.

Besides the fundamental interest of these processes, the generation of positrons finds application in surface science \cite{SL1988} through, for example, positron annihilation spectroscopy \cite{S1980_2,EVS06,TM13} and low-energy positron diffraction \cite{C02_2}, as well as in the study of their interactions with atoms and molecules \cite{SGB05,GYS10}. Positrons are also used to create antimatter (e.g., antihydrogen \cite{AAB02,GBO02,AAB10,GKK12}) and positronium \cite{CM07}). In these studies, slow positrons are commonly obtained from beta decay, decelerated through metallic moderators \cite{M1988}, and subsequently stored in different types of traps, from which they are extracted as low-energy, quasi-monochromatic pulses \cite{GKG97,GSM02,CDG06,NDS16}.

Direct positron generation from light would not require nuclear decay and could further leverage recent advances in optics to produce ultrashort photon pulses. However, the cross sections associated with the aforementioned processes are extremely small. As a possible avenue to increase the pair-production rate, we consider the replacement of free photons by confined optical modes in the hope that they alleviate the kinematic mismatch between the particles involved in BW scattering. In particular, surface polaritons, which are hybrids of light and polarization charges bound to material interfaces, can display short in-plane wavelengths compared with the free-space light wavelength. Actually, a broad suite of two-dimensional (2D) materials have recently been identified to sustain long-lived, strongly confined polaritons \cite{paper283,LCC17}, including plasmonic \cite{paper235,AND18,paper335}, phononic \cite{LDA18,CAC19}, and excitonic \cite{LCZ14} modes that cover a wide spectral range extending from mid-infrared frequencies \cite{paper235,AND18,LDA18,CAC19} to the visible domain \cite{paper335,LCZ14}. Specifically, modes bound to nanogaps \cite{paper156} feature large field confinement and enhancement (in vacuum regions) that boost light-mediated processes.

\begin{figure}
\begin{centering} \includegraphics[width=0.45\textwidth]{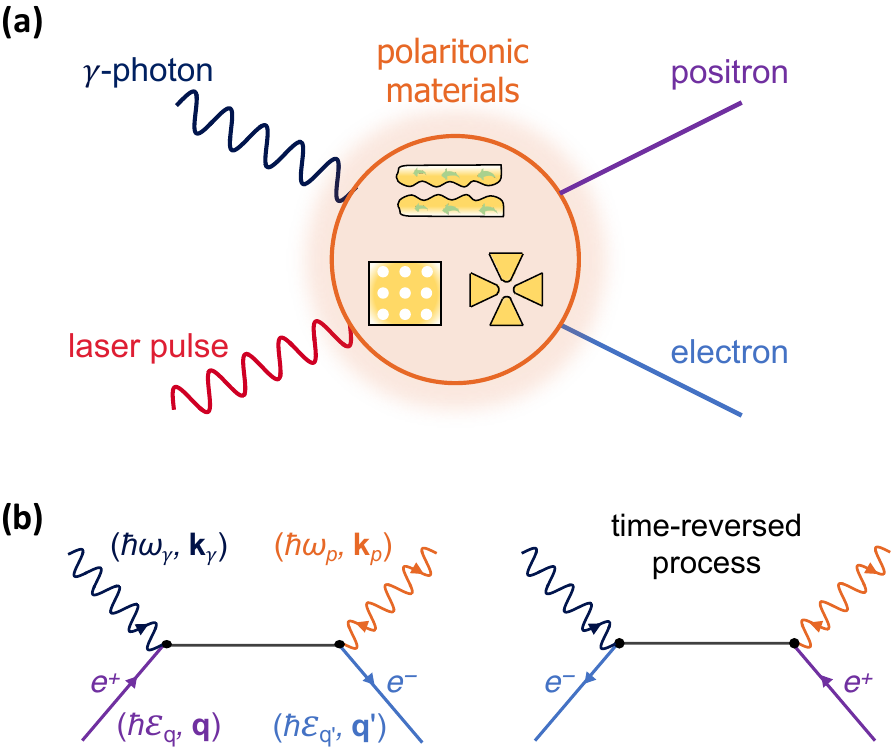} \par\end{centering}
\caption{{\bf Pair production by interaction of polaritons and $\gamma$-photons.} (a)~We consider polaritons supported by a material structure. Energetic $\gamma$-rays interact with the polaritons, giving rise to electron-positron pairs. (b)~Direct and time-reversed Feynman diagrams contributing to the investigated pair production. We indicate the energies and wave vectors of the polariton, the $\gamma$-photon, and the fermions by color-coordinated labels. Both polariton absorption and emission processes (double arrow) contribute to pair production.}
\label{Fig1}
\end{figure}

Here, we calculate the pair-production cross section associated with the annihilation of $\gamma$-ray photons ($\gamma$-photons) and confined polaritons, leading to a substantial enhancement compared to free-space BW scattering. Part of this enhancement relates to the spatial confinement of surface polaritons, as the lack of translational invariance enables pair production for $\gamma$-photon energies just above the $2\me c^2$ threshold (e.g., at the $^{60}$Co emission line $\hbar\og\sim1.17$~MeV combined with a polariton energy $\hbar\op$ of a few eV), in contrast to free-space BW scattering, for which visible-range photons need to be paired with GeV photons such as those existing in astrophysical processes \cite{R08}. For polaritonic nanogap modes confined in three dimensions, pairs are produced by $\gamma$-photon scattering in the gap vacuum region, where polariton-mediated positron emission is not affected by the background of other emission processes such as BH scattering. By demonstrating the advantages of using deeply confined light, our work inaugurates an avenue in the exploration of nanophotonic structures as a platform for high-energy physics.

\section{Pair production from the scattering of a polariton and a $\gamma$-photon}

Considering the general configuration illustrated in Fig.~\ref{Fig1}a, we study pair production by using the relativistic minimal coupling Hamiltonian \cite{JR1976,MS10}
\begin{align}
\hat{\mathcal{H}}_{\rm int}(t)=\frac{-1}{c} \int d^3\rb\; \hat{\jb}(\rb) \cdot \Ab(\rb,t),
\label{Hintmain}
\end{align}
where $\hat{\jb}(\rb)=\!-\ee c :\!\!\overline{\Psi}(\rb) \vec{\gamma}\hat{\Psi}(\rb)\!\!:$ is the fermionic current, $\Ab(\rb,t)$ is the classical vector potential associated with the polariton and photon fields, and we adopt a gauge with vanishing scalar potential. Here, $:\cdot:$ denotes normal product concerning electron and positron annihilation ($\hat{c}_{\qb,s}$ and $\hat{d}_{\qb,s}$, respectively) and creation ($\hat{c}^\dagger_{\qb,s}$ and $\hat{d}^\dagger_{\qb,s}$) operators (for fermions of momentum $\hbar\qb$, spin $s$, and energy $\hbar \varepsilon_{q}=c\sqrt{\me^2 c^2 + \hbar^2 q^2}$), and $\hat{\Psi}(\rb)=\sum_{\qb,s}\big(u_{\qb,s}\hat{c}_{\qb,s}\ee^{\ii\qb\cdot \rb}+v_{\qb,s}\hat{d}^\dagger_{\qb,s}\ee^{-\ii\qb\cdot\rb}\big)$ is the fermionic field operator, with $u_{\qb,s}$ ($v_{\qb,s}$) representing 4-component electron (positron) spinors.

We work in the continuous-wave regime and eventually normalize the resulting production rate to the number of polaritons and photons in the system. The vector potential is thus $\Ab(\rb,t)=-(\ii c/\op)\,\vec{\mathcal{E}}_{p}(\rb)\ee^{-\ii \op t}-(\ii c/\og)\,\mathcal{E}_{\gamma}\,\eh_j\,\ee^{\ii k_\gamma z-\ii \og t}+{\rm c.c.}$ (i.e., the sum of two monochromatic components), in which we consider two different polarizations $\eh_j=\xx$ or $\yy$ (with $j=1$ or 2) for the $\gamma$-ray field and take it to propagate along the $z$ direction with wave vector $\kbg=\zz\,\og/c$.

The production rate for a state $\hat{d}^\dagger_{\qb,s}\hat{c}^\dagger_{\qb',s'}\ket{0}$ comprising a positron (wave vector $\qb$, spin $s$) and an electron (wave vector $\qb'$, spin $s'$) is then calculated to the lowest (second) nonvanishing-order of time-dependent perturbation theory for the Hamiltonian in Eq.~(\ref{Hintmain}). This level of perturbation should be sufficient considering the low obtained cross sections (see below), while the renormalization group \cite{BN11,BJT15} could be used to account for nonperturbative corrections. Following a standard procedure detailed in Appendices~\ref{appendixA} and \ref{appendixB}, the positron-momentum-resolved pair-production cross section associated with polariton and $\gamma$-photon scattering is found to be
\begin{subequations}
\label{sgenMqqmain}
\begin{align}
\frac{d\sigma^{\rm pol}}{d\qb}&=\frac{\alpha^2c^5}{32\pi^4N_p\hbar\omega_\gamma\omega_p^2}
\int d^3\qb' \sum_\pm \sum_{ss'j} \label{sgenmain}\\
& \delta(\varepsilon_q+\varepsilon_{q'}-\omega_\gamma\pm\omega_p) \;
\bigg|\overline{u}_{\qb's'}\,\mathcal{M}^\pm_j(\qb',\qb)\,v_{\qb s}\bigg|^2, \nonumber
\end{align}
where $\alpha\approx1/137$ is the fine-structure constant, we average over $\gamma$-ray polarizations $j$, and a $4\times4$ matrix
\begin{align}
\mathcal{M}^\pm_j&(\qb',\qb)=\gamma^j\,G_F(\qb'-\kb_\gamma,\varepsilon_{q'}-\omega_\gamma)\,\vec{\gamma}\cdot\vec{\mathcal{E}}^{\pm}_{p,\qb+\qb'-\kbg}
\nonumber\\
&+\vec{\mathcal{E}}^{\pm}_{p,\qb+\qb'-\kbg}\cdot\vec{\gamma}\,G_F(\kb_\gamma-\qb,\omega_\gamma-\varepsilon_q)\,\gamma^j
\label{Mqqmain}
\end{align}
\end{subequations}
is defined in terms of the Dirac $\gamma$ matrices, the Feynman propagator \cite{MS10}
$G_F(\qb,\omega)=[\omega \gamma^0-c \vec{\gamma}\cdot \qb + (m_e c^2/\hbar)]/(\omega^2-\varepsilon_q^2+\ii 0^+)$, and the momentum representation of the polariton field $\vec{\mathcal{E}}^-_{p,\kb_p}=\int d^3 \rb~ \vec{\mathcal{E}}_{p}(\rb)\ee^{-\ii \kb_p \cdot \rb}$ and $\vec{\mathcal{E}}^+_{p,\kb_p}=\big(\vec{\mathcal{E}}^-_{p,-\kb_p}\big)^*$. The cross section in Eq.~(\ref{sgenmain}) is normalized per polariton and incident $\gamma$-photon, and in particular, the denominator in front of the integral contains the number of polaritons $N_p$ sustained by the $\vec{\mathcal{E}}_p(\rb)$ field (see Appendices~\ref{appendixA} and \ref{appendixB} for details).

Equations~(\ref{sgenMqqmain}) describe the annihilation of a $\gamma$-photon accompanied by the emission (upper signs) or absorption (lower signs) of a polariton, as indicated in the Feynman diagrams in Fig.~\ref{Fig1}b, where a finite range of wave vectors $\kb_p$ is generally involved due to spatial confinement. Incidentally, we note that boson emission is forbidden in free-space BW scattering, whereas it contributes to the present polariton-mediated pair-production process.

\begin{figure*}
\begin{centering} \includegraphics[width=1.0\textwidth]{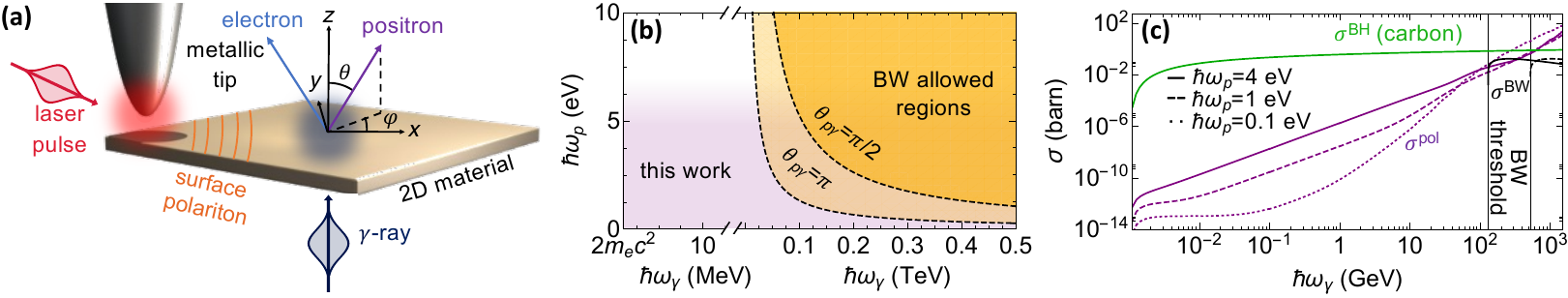} \par\end{centering}
\caption{{\bf Pair-production assisted by surface polaritons.} (a)~We consider surface modes excited in a 2D material by a coupling tip illuminated by laser pulses (red), while the $\gamma$-rays (dark gray) impinge normally to the surface. The positron emission direction $(\theta,\varphi)$ (purple) determines the electron direction (blue) by conservation of energy and in-plane momentum. (b)~Comparison between the regions allowed by energy-momentum conservation in either BW photon-photon scattering (yellow) and polariton-photon scattering under the configuration of Fig.~\ref{Fig1}a (purple) as a function of polariton/photon energies. The BW threshold $\hbar^2\op\og=2\me^2c^4/(1-\cos\theta_{p\gamma})$ \cite{JR1976} is indicated for a relative photon-photon angle $\theta_{p\gamma}$ of $\pi$ (absolute threshold) and $\pi/2$. (c)~Pair-production cross sections for polariton-photon scattering ($\sigma^{\rm pol}$, purple curves), BW scattering ($\sigma^{\rm BW}$ for $\theta_{p\gamma}=\pi/2$, black curves \cite{notepositron2}), and BH scattering by a carbon atom ($\sigma^{\rm BH}$, green curve \cite{notepositron3}). We consider different polariton energies (see legend) with a fixed $\kp=0.05$~nm$^{-1}$ in all cases. Solid vertical lines indicate the $\gamma$-photon BW threshold energies taken from the $\theta_{p\gamma}=\pi/2$ curve in (a).}
\label{Fig2}
\end{figure*}

\section{Pair production assisted by surface polaritons}

As an illustrative scenario, we consider surface polaritons (frequency $\op$, wave vector $\kb_p=\kp\xx$) launched on a 2D material ($z=0$ plane) by in-coupling a laser through a metallic tip (or, alternatively, a grating) (see Fig.~\ref{Fig2}a), producing a polariton field amplitude $\vec{\mathcal{E}}_{p}(\rb)\propto\big[i\kappa_p\xx-k_p{\rm sign}\{z\}\zz \big]\,\ee^{\ii\kp x-\kappa_p|z|}$ with $\kappa_p=\sqrt{k_p^2-\omega_p^2/c^2}$, where we neglect material losses, $\gamma$-ray screening, and finite-thickness effects. Parallel momentum conservation leads to $\qb'_{\parallel\pm}=-\qb_\parallel\mp\kb_p$ for the in-plane electron wave vector components, while energy conservation determines the electron energy $\varepsilon_{q'_\pm}=\og\mp\op-\varepsilon_{q}$ and out-of-plane wave vector $q'_{z\pm}=\sqrt{\varepsilon_{q'_\pm}^2/c^2-\me^2c^2/\hbar^2-q_{\parallel\pm}^{\prime\,2}}$, subject to the threshold-energy conditions $\varepsilon_{q'_\pm}^2>\me^2c^4/\hbar^2-c^2q_{\parallel\pm}^{\prime\,2}$ and $\og>\pm\op+\varepsilon_{q}$. Calculating the Fourier transform of the surface polariton field and inserting it into Eqs.~(\ref{sgenMqqmain}), we find (see Appendix~\ref{appendixD})
\begin{align}
\frac{d\sigma^{\rm pol}}{d\qb}
=&\frac{\alpha^2c^3\kappa_p}{\pi\,\omega_p\omega_\gamma k_p^2}
\sum_\pm
\frac{\varepsilon_{q'_\pm}}{q'_{z\pm}}
\label{sigmapolmain}\\
&\times\sum_{ss'j\mu} \;
\Big|\overline{u}_{\qb'_{\mu\pm},s'}\mathcal{N}^\pm_j(\qb'_{\mu\pm},\qb)\,v_{\qb s}\Big|^2, \nonumber
\end{align}
where $\qb'_{\mu\pm}=\qb'_{\parallel\pm}+\mu q'_{z\pm}\zz$ is the electron wave vector for upward ($\mu=1$) and downward ($\mu=-1$) emission contributions, while $\mathcal{N}^\pm_j(\qb',\qb)$ is given by Eq.~(\ref{Mqqmain}) with $\vec{\mathcal{E}}^{\pm}_{p,\qb+\qb'-\kbg}$ replaced by $\fb_{\pm(k_{\gamma z}-q_z-q'_z)}$. Here, $\fb_{k_z}=(\kappa_p^2\,\xx+\kp k_z\,\zz)/(\kappa_p^2+ k_z^2)$ encapsulates the out-of-plane momentum distribution of the polariton field.

An immediate consequence of out-of-plane symmetry breaking is that the allowed kinematical space for which we obtain nonzero pair-production cross sections extends down to the infrared polariton regime even when using $\gamma$-photons just above the absolute energy threshold $\gtrsim2\me c^2\approx1.02$~MeV (Fig.~\ref{Fig2}b). In contrast, BW scattering with one of the photons in the optical regime requires the other photon to have energy exceeding $\sim0.1$~TeV, which explains why free-space pair production has traditionally been observed only in its nonlinear version, where the energy-momentum mismatch is overcome by engaging a high number of photon exchanges \cite{R1961,R1985}.

In Fig.~\ref{Fig2}c, we show that, for low-energy polaritons/photons (up to a few eV), the momentum-integrated polariton-assisted pair-production cross section $\sigma^{\rm pol}=\int d^3\qb\;(d\sigma^{\rm pol}/d\qb)$, with $d\sigma^{\rm pol}/d\qb$ given by Eq.~(\ref{sigmapolmain}), takes substantial values at $\gamma$-photon energies far below the BW kinematical threshold (vertical solid lines). In addition, $\sigma^{\rm pol}$ is consistently several orders of magnitude higher than the BW cross section up to $\gamma$-photon energies in the TeV regime. Part of this enhancement can be attributed to the effect of spatial compression of polaritons relative to free-space photons.

Upon numerical examination of Eq.~(\ref{sigmapolmain}), we find positron emission to be dominated by contributions associated with an equal partition of kinetic energy between the two fermions, both for near-threshold (Fig.~\ref{FigS2}) and GeV (Fig.~\ref{FigS3}) emission, also displaying sharp angular profiles peaked around the forward direction defined by the $\gamma$-ray.

\begin{figure}
\begin{centering} \includegraphics[width=0.4\textwidth]{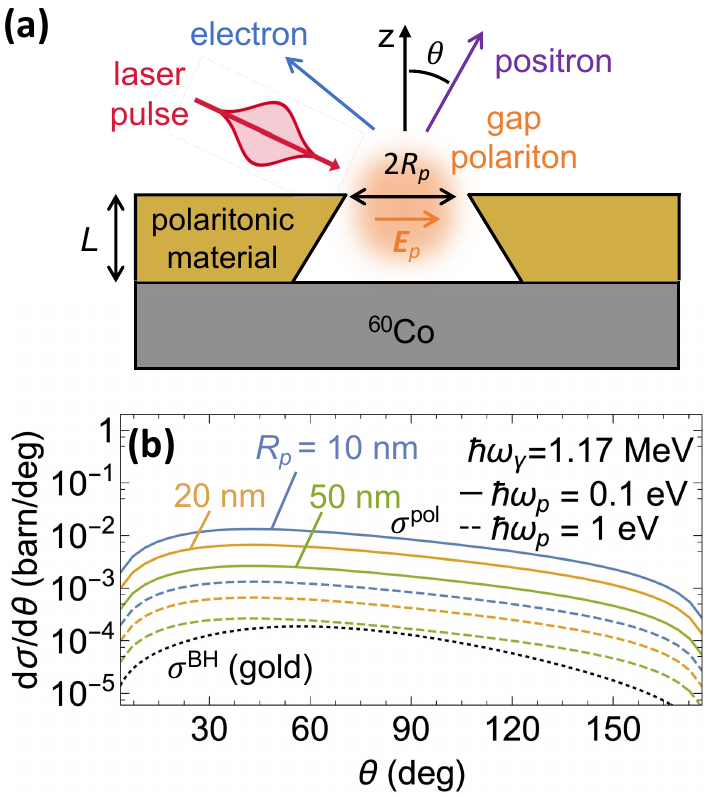} \par\end{centering}
\caption{{\bf Pair-production from a gap polariton.} (a)~Sketch of the geometry under consideration, in which pairs are produced by $\gamma$-photons traversing a gap polariton. The latter can be excited by a laser pulse and is taken to have energy $\omega_p$ and uniform field $\vec{\mathcal{E}}_p$ confined to a spherical region of radius $R_p$ (flanked by a polaritonic material). (b)~Differential cross section as a function of polar angle for polariton-assisted positron emission under the configuration in (a) (colored curves for different values of $R_p$ and $\omega_p$, as indicated by labels), compared with the BH cross section for a gold atom \cite{notepositron3}. We consider 1.17\,MeV $\gamma$-photons in all cases.}
\label{Fig3}
\end{figure}

Unfortunately, the emission arising from scattering by surface polaritons is orders of magnitude smaller than that associated with BH scattering by the polaritonic material, as revealed by comparing their respective cross sections normalized per polariton and per atom (Fig.~\ref{Fig3}c). For example, for 1.17~MeV $\gamma$-photons traversing a highly doped graphene monolayer that supports 1~eV plasmons, the ratio between the emission from these two mechanisms is $(n_p\sigma^{\rm pol})/(n_C\sigma^{\rm BH})$, where $n_p$ is the plasmon surface density, $n_C\sim40/$nm$^2$ is the carbon atom density, and we have $\sigma^{\rm pol}\sim10^{-13}$~barn ($1$~barn\,$=10^{-24}$ cm$^2$) and $\sigma^{\rm BH}\sim10^{-4}$~barn \cite{M1968,notepositron3}. For the two signals to be comparable in magnitude, an unrealistically large plasmon density $n_p>10^{10}/$nm$^2$ would be required.

\section{Threshold pair-production assisted by gap polaritons}

To reduce the effect of the BH background, we study pair production by scattering of $\gamma$-photons and gap polaritons (Fig. \ref{Fig3}a). Besides the emission enhancement expected from the breaking of translational invariance in all directions, positrons produced by gap polaritons and $\gamma$-photons arise from the vacuum gap region, where no BH signal is generated, thus facilitating the identification of a polariton-assisted pair-production signal (see further discussion below). For simplicity, we consider a polariton field $\vec{\mathcal{E}}_p(\rb)=E_p\,\xx\,\Theta(R_p-r)$ of uniform amplitude $E_p$ polarized along $x$ and confined within a sphere of radius $R_p$. Inserting this field in Eqs.~(\ref{sgenMqqmain}), we obtain a semi-analytical expression (see Appendix~\ref{appendixC}) from which the results presented in Fig.~\ref{Fig3}b are computed for different polariton sizes $R_p$ and energies $\hbar \omega_p$ after integrating over the azimuthal angle of positron emission. The differential cross section normalized per polariton and $\gamma$-photon exhibits a monotonic increase with decreasing $R_p$ and $\omega_p$ as well as a smooth dependence on polar angle $\theta$.

Once more, we need to compare polariton-driven pair production to the background BH positron signal (i.e., $\gamma$-ray scattering by the nuclei of the polaritonic material). The complete suppression of BH scattering from the vacuum gap region could be leveraged by selecting positrons originating only in that region through the use of charged-particle optics elements (i.e., a positron analog of electron optics in an electron microscope), such that only positrons coming from the gap are collected, similarly to how photoemission electron microscopes collect electrons emitted within specimen regions spanning just a few nanometers \cite{MYH20}.

Even without resorting to positron microscopy, we argue next that spatial confinement in gap polaritons leads to a discernible positron emission signal under laser pulse irradiation when compared to the BH background, as the cross section per polariton undergoes an increase by several orders of magnitude when moving from confinement in one direction (surface polaritons, Fig.~\ref{Fig2}c) to full three-dimensional trapping (gap polaritons, Fig.~\ref{Fig3}b). For concreteness, we focus on low-energy ($\hbar\omega_p=0.1$\,eV) gap plasmons confined to an opening in a gold film with an effective mode volume assimilated to a sphere of radius $R_p=50\,$nm. These parameters can be obtained by engineering the morphology of the metal gap \cite{paper156}. In practice, we envision an array of gaps such that the openings span a fraction $\eta$ of the film surface. Under illumination with a laser peak amplitude of $10^8$\,V/m (a typical value below the damage threshold when using ultrafast pulses \cite{FES15}) and a realistic polaritonic field enhancement of $10^2$ (i.e., $E_p\sim10^{10}$~V/m), we have a number of polaritons $N_p\approx E_p^2R_p^3/3\hbar \omega_p\sim3\times10^7$ per gap (i.e., a surface polariton density $n_p=\eta N_p/\pi R_p^2\sim4\,\eta\times10^3/$nm$^2$; see Appendix~\ref{appendixD}), and therefore, the fraction of positrons generated per $\gamma$-photon is $n_p\sigma^{\rm pol}\sim\eta\times10^{-7}$, where we take $\sigma^{\rm pol}\sim0.25\,$barn for the pair-production cross section per polariton (see Fig.~\ref{Fig3}b).

This fraction has to be compared to that of positrons associated with the BH mechanism. For a gold film of thickness $L=100$~nm ($\ll$ positron escape depth \cite{notepositron6}), as commonly employed in plasmonic studies, we combine the BH cross section for a gold atom at 1.17~MeV $\gamma$-photon energy ($\sigma^{\rm BH}\approx16~$mbarn; see Fig.~\ref{Fig3}b) together with the volume per gold atom $\mathcal{V}\approx17.0\,{\AA}^3$ (i.e., a surface gold atom density $n_{\rm Au}=L/\mathcal{V}\sim3\times10^5/$nm$^2$), to compute the fraction of BH positrons per incident photon, $n_{\rm Au}\sigma^{\rm BH}\sim10^{-8}$. Under these conditions, the ratio of polariton-assisted emission to BH emission is $\sim10\,\eta$. For a realistic value of the opening fraction $\eta\sim10\%$, the noted ratio becomes $\sim1$, and therefore, polaritons and BH scattering are comparable in magnitude.

We remark that this estimate assumes a synchronized detection, such that the signal is only collected within the duration of the optical pulses needed to sustain a large number of polaritons in the system. For example, with 1\,g of $^{60}$Co, we have $\sim10^2$ $\gamma$-photons overlapping with the duration of a 1\,ps laser pulse (see Appendix~\ref{appendixD}), which lead to the emission of $n_p\sigma^{\rm pol}\sim10^{-6}$ positrons per pulse, half of them produced by polariton-assisted scattering. We thus predict a measurable signal when employing a high-repetition ($\sim10^8$~Hz) pulsed laser.

As an alternative geometry, one could rely on polaritons confined to nanoparticles (e.g., gold colloids \cite{paper300}) of similar size as the gaps considered above and dispersed on a thin film (e.g., monolayer graphene), leading to similar estimates for the positron production yield and even higher ratios of polariton-driven to BH positron emission because of the reduction in polaritonic material volume.

\section{Conclusion}

In conclusion, we advocate for the use of optical excitations confined to nanostructured materials in combination with $\gamma$-rays as a way of producing electron-positron pairs with higher efficiency than free-space BW scattering and requiring substantially lower photon energies. The breaking of translational invariance is responsible for the latter, whereas the spatial compression of the optical fields associated with surface polaritons facilitates the coupling to high-momentum products (the fermions). The proposed mechanism is still orders of magnitude weaker than BH scattering when the pairs are produced by $\gamma$-photons traversing polariton-supporting materials (e.g., graphene and planar waveguides). In contrast, we argue that gap polaritons confined to vacuum regions flanked by such materials can circumvent this problem by, for example, collecting positrons created at the gap using a positron microscope. In addition, when synchronizing positron detection with exposure of gap polaritons to ultrafast laser pulses, we show that the BH background becomes comparatively small thanks to the enhancement in the emission associated with strong polariton spatial confinement in all three dimensions. We remark that these conclusions are drawn from the study of positron emission produced by near-the-threshold $\gamma$-photons, such as those from $^{60}$Co.

Besides its fundamental interest, the proposed mechanism for polariton-driven positron emission opens exciting possibilities that are not accessible to other types of positron sources, such as the generation of positron pulses with ultrafast durations inherited from the incident laser pulses. We envision the spatiotemporal modulation of the positron wave functions by shaping the employed laser field (e.g., to create chiral positron beams or two-pulse positron states). Moreover, the localized nature of the emission from regions in which strongly confined and intensity-enhanced polaritons are sustained renders this mechanism appealing for applications that demand spatially confined positron sources. Beyond these potential uses, antimatter production from collective optical excitations bears intrinsic interest as an example of a nanophotonics approach to high-energy physics.

\begin{widetext}
\section*{APPENDIX}
\appendix


\renewcommand{\thefigure}{S\arabic{figure}} 
\renewcommand{\thetable}{S\arabic{table}} 

\section{Pair-production matrix elements for polychromatic classical electromagnetic fields}
\label{appendixA}
\renewcommand{\theequation}{A\arabic{equation}} 

To be tutorial for researchers in fields including nanophotonics and quantum optics, we provide a detailed derivation of the pair-production cross section based on standard second-order perturbation theory. We supplement this calculation in Sec.~\ref{general} below by following the quantum field theory formalism of quantum electrodynamics (QED) \cite{MS10}, more commonly used in the high-energy physics community.

\subsection{Time-dependent perturbation theory}

Before specifying the calculation for pair production, we review a general formalism for perturbation theory. A complete set of eigenstates $\ket{j}$ of the unperturbed Hamiltonian $\mathcal{H}_0$ is considered, characterized by energies $\hbar\varepsilon_j$ and, therefore, satisfying $\mathcal{H}_0\ket{j}=\hbar\varepsilon_j\ket{j}$. Such eigenstates will later be identified with electron and positron number states. We also introduce the perturbation produced by a time-dependent Hamiltonian $\mathcal{H}_{\rm int}(t)$ of matrix elements $\bra{j}\mathcal{H}_{\rm int}(t)\ket{j'}=\sum_i\big(V_{jj'}^i\ee^{-\ii\omega_it}+V_{j'j}^{i*}\ee^{\ii\omega_it}\big)$, where the $i$ sum runs over components evolving with frequencies $\omega_i$ (corresponding to the interaction with classical polariton and $\gamma$-ray fields in the present study). We now expand the time-dependent state of the system as $\ket{\psi(t)}=\sum_j\alpha_j(t)\,\ee^{-\ii\varepsilon_jt}\ket{j}$, whose evolution is ruled by the Schr\"odinger equation $[\mathcal{H}_0+\mathcal{H}_{\rm int}(t)]\ket{\psi(t)}=\ii\hbar\partial_t\ket{\psi(t)}$ or, equivalently, the equation of motion
\begin{align}
\dot{\alpha}_{j}(t)=-\frac{\ii}{\hbar}\sum_{ij'}\big(V_{jj'}^i\,\ee^{-\ii\omega_it}+V_{j'j}^{i*}\,\ee^{\ii\omega_it}\big)\;\ee^{\ii\varepsilon_{jj'}t}\;\alpha_{j'}(t) \nonumber
\end{align}
for the expansion coefficients, where we use the notation $\varepsilon_{jj'}=\varepsilon_{j}-\varepsilon_{j'}$.

Starting from the nondegenerate ground state $\ket{j=0}$ at $t=-\infty$ (the fermionic vacuum in this work), we write the perturbation series $\alpha_j(t)=\sum_n\alpha_j^{(n)}(t)$, where $n$ is the order of interaction and $\alpha_j^{(0)}=\delta_{j0}$ describes the unperturbed state. The first- and second-order terms can be readily obtained upon direct integration as
\begin{align}
\alpha_j^{(1)}(t)=-\frac{1}{\hbar}\sum_{i}\bigg[&
\frac{V_{j0}^i\,\ee^{\ii(\varepsilon_{j0}-\omega_i-\ii\delta)t}}{\varepsilon_{j0}-\omega_i-\ii\delta}
+\frac{V_{0j}^{i*}\,\ee^{\ii(\varepsilon_{j0}+\omega_i-\ii\delta)t}}{\varepsilon_{j0}+\omega_i-\ii\delta}
\bigg], \nonumber\\
\alpha_{j}^{(2)}(t)=\frac{1}{\hbar^2}\sum_{ii'j'}\bigg[&
\frac{\ee^{\ii(\varepsilon_{j0}-\omega_i-\omega_{i'}-\ii\delta)t}}{(\varepsilon_{j0}-\omega_i-\omega_{i'}-\ii\delta)}
\frac{V_{jj'}^iV_{j'0}^{i'}}{(\varepsilon_{j'0}-\omega_{i'}-\ii\delta)}
+\frac{\ee^{\ii(\varepsilon_{j0}+\omega_i+\omega_{i'}-\ii\delta)t}}{(\varepsilon_{j0}+\omega_i+\omega_{i'}-\ii\delta)}
\frac{V_{j'j}^{i*}V_{0j'}^{i'*}}{(\varepsilon_{j'0}+\omega_{i'}-\ii\delta)} \nonumber\\
+&\frac{\ee^{\ii(\varepsilon_{j0}-\omega_i+\omega_{i'}-\ii\delta)t}}{(\varepsilon_{j0}-\omega_i+\omega_{i'}-\ii\delta)}\bigg(
\frac{V_{jj'}^iV_{0j'}^{i'*}}{\varepsilon_{j'0}+\omega_{i'}-\ii\delta}
+\frac{V_{j'j}^{i'*}V_{j'0}^{i}}{\varepsilon_{j'0}-\omega_i-\ii\delta}\bigg)
\bigg], \nonumber
\end{align}
where $\delta$ is a positive infinitesimal introduced to adiabatically switch on the interaction. The transition rate at a finite time $t$ is then given by $\Gamma_{0\to j}=\lim_{\delta\to0^+}d|\alpha_j(t)|^2/dt$. In particular, within first-order perturbation theory (i.e., retaining terms up to order $n=1$ in the perturbation series), we find
\begin{subequations}
\label{G12}
\begin{align}
\Gamma_{0\to j}^{(1)}=\frac{2\pi}{\hbar^2}\sum_{i}
\big|V_{j0}^i\big|^2\,\delta(\varepsilon_{j0}-\omega_i).
\label{G1}
\end{align}
If $\Gamma_{0\to j}^{(1)}$ vanishes, the next leading contribution to the transition rate comes from $\alpha_j^{(2)}(t)$, which yields
\begin{align}
\Gamma_{0\to j}^{(2)}=\frac{2\pi}{\hbar^4}\sum_{i}\Bigg[
&\bigg|\sum_{j'}\frac{V_{jj'}^iV_{j'0}^i}{\varepsilon_{j'0}-\omega_i-\ii\delta}
\bigg|^2\,\delta(\varepsilon_{j0}-2\omega_i) \nonumber\\
+&\sum_{i'<i} \bigg|\sum_{j'}\bigg(\frac{V_{jj'}^iV_{j'0}^{i'}}{\varepsilon_{j'0}-\omega_{i'}-\ii\delta}
+\frac{V_{jj'}^{i'}V_{j'0}^i}{\varepsilon_{j'0}-\omega_i-\ii\delta}\bigg)\bigg|^2\,\delta(\varepsilon_{j0}-\omega_i-\omega_{i'}) \nonumber\\
+&\sum_{i'}\bigg|\sum_{j'}\bigg(
\frac{V_{jj'}^iV_{0j'}^{i'*}}{\varepsilon_{j'0}+\omega_{i'}-\ii\delta}
+\frac{V_{j'j}^{i'*}V_{j'0}^{i}}{\varepsilon_{j'0}-\omega_i-\ii\delta}
\bigg)\bigg|^2\,\delta(\varepsilon_{j0}-\omega_i+\omega_{i'}) \Bigg].
\label{G2}
\end{align}
\end{subequations}
In the derivation of Eqs.~(\ref{G12}), we have used the fact that $\varepsilon_j>\varepsilon_0$ and assumed nondegenerate frequencies $\omega_i$ and frequency differences $\omega_i-\omega_{i'}$ for $i\neq i'$. In addition, the contribution arising from terms containing two negative energies vanishes because they cannot conserve energy.

\subsection{QED Hamiltonian and matrix elements}
\label{perturbation}

We study pair production produced by a classical electromagnetic field that is described through the vector potential $\Ab(\rb,t)$ in the temporal gauge (i.e., with a vanishing scalar potential \cite{notepositron,JR1976}). We adopt the minimal-coupling relativistic QED Hamiltonian in the Schr\"odinger picture \cite{MS10}
\begin{align}
\hat{\mathcal{H}}_{\rm int}(t)= -\frac{1}{c} \int d^3\rb\;\hat{\jb}(\rb)\cdot \Ab(\rb,t),\label{hint}
\end{align}
where $\hat{\jb}(\rb,t) = -\ee c :\overline{\Psi}(\rb) \vec{\gamma}\hat{\Psi}(\rb):$ is the current operator, we define $\overline{\Psi}=\hat{\Psi}^\dagger \gamma^0$, and the notation $:\boldsymbol{\cdot}:$ is used to indicate normal product acting on the fermionic field operators $\hat{\Psi}(\rb)$ and $\hat{\Psi}^\dagger(\rb)$. Here, $\vec{\gamma}$ and $\gamma^0$ \cite{MS10} are the spatial and temporal Dirac matrices. The field operator is then expanded as
\begin{align}
\hat{\Psi}(\rb)=\frac{1}{\sqrt{V}}\sum_{\qb,s}\left(u_{\qb,s}\hat{c}_{\qb,s}\ee^{\ii\qb\cdot \rb}+v_{\qb,s}\hat{d}^\dagger_{\qb,s}\ee^{-\ii\qb\cdot\rb}\right),
\nonumber
\end{align}
where $V$ is the normalization volume and we introduce the anticommuting annihilation operators $\hat{c}_{\qb,s}$ and $\hat{d}_{\qb,s}$ and the corresponding creation operators $\hat{c}^\dagger_{\qb,s}$ and $\hat{d}^\dagger_{\qb,s}$ for electron and positron plane waves of wave vector $\qb$ and spin $s$. The associated 4-component electron and positron spinors $u_{\qb,s}$ and $v_{\qb,s}$ are chosen to satisfy the equations 
\begin{subequations}
\label{Dirac}
\begin{align}
(\hbar c\,\vec{\gamma}\cdot\qb +m_e c^2\mathcal{I}_4)\,u_{\qb,s}=\hbar \varepsilon_q \gamma^0 u_{\qb,s},\\
(\hbar c\,\vec{\gamma}\cdot\qb -m_e c^2\mathcal{I}_4)\,v_{\qb,s}=\hbar \varepsilon_q \gamma^0 v_{\qb,s},
\end{align}
\end{subequations}
subject to the orthonormalization conditions $u_{\qb,s}^\dagger u_{\qb,s'}=\delta_{s,s'}$, $v_{\qb,s}^\dagger v_{\qb,s'}=\delta_{s,s'}$ and $u_{\qb,s}^\dagger v_{-\qb,s'}=0$. Here, $m_e$ is the electron/positron mass, $\hbar\varepsilon_q = c \sqrt{\me^2 c^2+\hbar^2 q^2}$ is the relativistic particle energy, and $\mathcal{I}_4$ is the $4\times4$ identity matrix.

The current operator takes the explicit form
\begin{align}
\hat{\jb}(\rb)=-\frac{ec}{V}\sum_{\qb\qb'}\sum_{ss'}\big[
&\hat{c}^\dagger_{\qb s}\hat{c}_{\qb's'}\,\ee^{\ii(\qb'-\qb)\cdot\rb}\,\overline{u}_{\qb s}\vec{\gamma}\,u_{\qb's'}
-\hat{d}^\dagger_{\qb s}\hat{d}_{\qb's'}\,\ee^{\ii(\qb'-\qb)\cdot\rb}\,\overline{v}_{\qb's'}\vec{\gamma}\,v_{\qb s} \nonumber\\
-&\hat{d}^\dagger_{\qb s}\hat{c}^\dagger_{\qb's'}\,\ee^{-\ii(\qb+\qb')\cdot\rb}\,\overline{u}_{\qb's'}\vec{\gamma}\,v_{\qb s}
-\hat{c}_{\qb's'}\hat{d}_{\qb s}\,\ee^{\ii(\qb+\qb')\cdot\rb}\,\overline{v}_{\qb s}\vec{\gamma}\,u_{\qb's'}
\big], \label{jr}
\end{align}
where the first two terms describe electron and positron scattering, while the remaining two terms stand for pair creation and annihilation, respectively. In addition, the electromagnetic field is taken to consist of monochromatic components of frequencies $\omega_i$, such that the vector potential can be written
\begin{align}
\Ab(\rb,t)=-\ii\,c\sum_{i}\frac{1}{\omega_i} \,\vec{\mathcal{E}}_i(\rb)\ee^{-\ii \omega_i t}+{\rm c.c.}, \label{Art}
\end{align}
where $\vec{\mathcal{E}}_i(\rb)$ are the time-independent amplitudes of the field components.

We are now prepared to evaluate the matrix elements of the interaction Hamiltonian in Eq.~(\ref{hint}), entering the rates in Eqs.~(\ref{G12}) with the $j$ labels running over electron-positron pairs. We thus multiplex $\ket{j}$ as $\ket{p\qb s,e\qb's'}=\hat{d}^\dagger_{\qb s}\hat{c}^\dagger_{\qb's'}\ket{0}$, where $e$ and $p$ refer to electrons and positrons, respectively. Using Eqs.~(\ref{hint}), (\ref{jr}), and (\ref{Art}), we find
\begin{align}
V^i_{p\qb s,e\qb's';p\qb s,e\qb''s''}=-&\frac{\ii ec}{V\omega_i}\,\overline{u}_{\qb's'}\,\vec{\gamma}\cdot\vec{\mathcal{E}}_{i,\qb'-\qb''}\,u_{\qb''s''}, & \text{electron scattering} \nonumber\\
V^i_{p\qb's',e\qb s;p\qb''s'',e\qb s}=+&\frac{\ii ec}{V\omega_i}\,\overline{v}_{\qb''s''}\,\vec{\gamma}\cdot\vec{\mathcal{E}}_{i,\qb'-\qb''}\,v_{\qb's'}, & \text{positron scattering} \nonumber\\
V^i_{p\qb s,e\qb's';0}=+&\frac{\ii ec}{V\omega_i}\,\overline{u}_{\qb's'}\,\vec{\gamma}\cdot\vec{\mathcal{E}}_{i,\qb+\qb'}\,v_{\qb s}, & \text{pair creation} \nonumber\\
V^i_{0;p\qb s,e\qb's'}=+&\frac{\ii ec}{V\omega_i}\,\overline{v}_{\qb s}\,\vec{\gamma}\cdot\vec{\mathcal{E}}_{i,-\qb-\qb'}\,u_{\qb's'}, & \text{pair annihilation} \nonumber
\end{align}
where
\begin{align}
\vec{\mathcal{E}}_{i,\kb}=\int d^3 \rb~ \vec{\mathcal{E}}_{i}(\rb)\ee^{-\ii \kb \cdot \rb}
\label{Eik}
\end{align}
is the Fourier transform of the field amplitudes, which imposes momentum conservation.

\subsection{Pair-production rate in second-order perturbation theory}

Pair creation by a single photon is kinematically forbidden (i.e., $\Gamma_{0\to j}^{(1)}=0$), and thus, we need to go to the second order in the light-matter interaction, for which we use Eq.~(\ref{G2}). In the evaluation of $\Gamma_{0\to j}^{(2)}$, it is convenient to analytically carry out the internal sums over $j'$. Taking the final product as $j\to p\qb s,e\qb's'$, the sums in the first and second lines of Eq.~(\ref{G2}) can be evaluated as follows:
\begin{align}
&\sum_{j'}\bigg(\frac{V_{jj'}^iV_{j'0}^{i'}}{\varepsilon_{j'0}-\omega_{i'}-\ii\delta}
+\frac{V_{jj'}^{i'}V_{j'0}^i}{\varepsilon_{j'0}-\omega_i-\ii\delta}\bigg) \nonumber\\
&=\sum_{\qb''s''}\bigg[
\frac{V_{p\qb s,e\qb's';p\qb s,e\qb''s''}^iV_{p\qb s,e\qb''s'';0}^{i'}}{\varepsilon_{q''}+\varepsilon_{q}-\omega_{i'}-\ii\delta}
+\frac{V_{p\qb s,e\qb's';p\qb''s'',e\qb's'}^iV_{p\qb''s'',e\qb's';0}^{i'}}{\varepsilon_{q'}+\varepsilon_{q''}-\omega_{i'}-\ii\delta} \nonumber \\
&\quad\quad\quad+\frac{V_{p\qb s,e\qb's';p\qb s,e\qb''s''}^{i'}V_{p\qb s,e\qb''s'';0}^i}{\varepsilon_{q''}+\varepsilon_{q}-\omega_i-\ii\delta}
+\frac{V_{p\qb s,e\qb's';p\qb''s'',e\qb's'}^{i'}V_{p\qb''s'',e\qb's';0}^i}{\varepsilon_{q'}+\varepsilon_{q''}-\omega_i-\ii\delta}
\bigg] \nonumber \\
&=\frac{e^2c^2}{V^2\omega_i\omega_{i'}}\sum_{\qb''s''}\bigg[
\frac{\overline{u}_{\qb's'}\,\vec{\gamma}\cdot\vec{\mathcal{E}}_{i,\qb'-\qb''}\,u_{\qb''s''}
\,\overline{u}_{\qb''s''}\,\vec{\gamma}\cdot\vec{\mathcal{E}}_{i',\qb+\qb''}\,v_{\qb s}
}{\varepsilon_{q''}-(\varepsilon_{q'}-\omega_i)-\ii\delta}
-\frac{\overline{v}_{\qb''s''}\,\vec{\gamma}\cdot\vec{\mathcal{E}}_{i,\qb-\qb''}\,v_{\qb s}
\,\overline{u}_{\qb's'}\,\vec{\gamma}\cdot\vec{\mathcal{E}}_{i',\qb'+\qb''}\,v_{\qb''s''}
}{\varepsilon_{q''}+(\varepsilon_{q'}-\omega_{i'})-\ii\delta} \nonumber\\
&\quad\quad\quad\quad\quad\quad\quad\;+\frac{\overline{u}_{\qb's'}\,\vec{\gamma}\cdot\vec{\mathcal{E}}_{i',\qb'-\qb''}\,u_{\qb''s''}
\,\overline{u}_{\qb''s''}\,\vec{\gamma}\cdot\vec{\mathcal{E}}_{i,\qb+\qb''}\,v_{\qb s}
}{\varepsilon_{q''}-(\varepsilon_{q'}-\omega_{i'})-\ii\delta}
-\frac{\overline{v}_{\qb''s''}\,\vec{\gamma}\cdot\vec{\mathcal{E}}_{i',\qb-\qb''}\,v_{\qb s}
\,\overline{u}_{\qb's'}\,\vec{\gamma}\cdot\vec{\mathcal{E}}_{i,\qb'+\qb''}\,v_{\qb''s''}
}{\varepsilon_{q''}+(\varepsilon_{q'}-\omega_i)-\ii\delta}
\bigg] \nonumber \\
&=-\frac{e^2c^2}{V^2\omega_i\omega_{i'}}\overline{u}_{\qb's'}\,\vec{\gamma}\cdot\sum_{\qb''}\Big[
\vec{\mathcal{E}}_{i,\qb'-\qb''}
\,G_F(\qb'',\varepsilon_{q'}-\omega_i)\,\vec{\mathcal{E}}_{i',\qb+\qb''}
+\vec{\mathcal{E}}_{i',\qb'-\qb''}\,
G_F(\qb'',\varepsilon_{q'}-\omega_{i'})\,\vec{\mathcal{E}}_{i,\qb+\qb''}
\Big]\cdot\vec{\gamma}\,v_{\qb s}, \label{sumjp1}
\end{align}
where
\begin{align}
G_F(\qb,\omega)&=-\sum_s\bigg(\frac{u_{\qb s}\,\overline{u}_{\qb s}}{\varepsilon_q-\omega-\ii\delta}
-\frac{v_{-\qb s}\,\overline{v}_{-\qb s}}{\varepsilon_q+\omega-\ii\delta}\bigg) \nonumber\\
&=\frac{\omega \gamma^0-c \vec{\gamma}\cdot \qb + (m_e c^2/\hbar)\, \mathcal{I}_4}{\omega^2-\varepsilon_q^2+\ii 0^+}. \label{GFF}
\end{align}
is the so-called Feynman propagator \cite{MS10}. In the derivation of this result, we have invoked energy conservation (i.e., the condition $\varepsilon_{q}+\varepsilon_{q'}=\omega_i+\omega_{i'}$ imposed by the $\delta$-function in Eq.~(\ref{G2})) and changed $\qb''\to-\qb''$ in the positron-scattering terms. In addition, the second line of Eq.~(\ref{GFF}) is obtained from the first one by first combining the two fractions and then using Eqs.~(\ref{Dirac}) to eliminate $\varepsilon_q$ in the numerator, applying the completeness relation $\sum_s\big(u_{\qb s}\,u^\dagger_{\qb s}+v_{-\qb s}\,v^\dagger_{-\qb s}\big)=\mathcal{I}_4$, and taking the $\delta\to0^+$ limit. Following a similar procedure and making use of the identity $(\overline{u}\,\vec{\gamma}\,v)^*=-\overline{v}\,\vec{\gamma}\,u$, we find
\begin{align}
&\sum_{j'}\bigg(\frac{V_{jj'}^iV_{0j'}^{i'*}}{\varepsilon_{j'0}+\omega_{i'}-\ii\delta}
+\frac{V_{j'j}^{i'*}V_{j'0}^i}{\varepsilon_{j'0}-\omega_i-\ii\delta}\bigg) \nonumber\\
&=\frac{e^2c^2}{V^2\omega_i\omega_{i'}}\overline{u}_{\qb's'}\,\vec{\gamma}\cdot\sum_{\qb''}\Big[
\vec{\mathcal{E}}_{i,\qb'-\qb''}
\,G_F(\qb'',\varepsilon_{q'}-\omega_i)\,\vec{\mathcal{E}}^*_{i',-\qb-\qb''}
+\vec{\mathcal{E}}^*_{i',\qb''-\qb'}\,
G_F(\qb'',\varepsilon_{q'}+\omega_{i'})\,\vec{\mathcal{E}}_{i,\qb+\qb''}
\Big]\cdot\vec{\gamma}\,v_{\qb s} \label{sumjp2}
\end{align}
for the $j'$ sum in the third line of Eq.~(\ref{G2}).

Finally, using Eqs.~(\ref{sumjp1}) and (\ref{sumjp2}) in Eq.~(\ref{G2}) and ignoring contributions from two photons of the same frequency (because we are interested in polariton and $\gamma$-ray scattering), we find the second-order pair-production rate
\begin{align}
\Gamma_{p\qb s,e\qb's'}^{(2)}=&\frac{2\pi e^4c^4}{V^4 \hbar^4}\sum_{ii'}{}^{\prime}\frac{1}{\omega_i^2\omega_{i'}^2}\sum_\pm\delta(\varepsilon_q+\varepsilon_{q'}-\omega_i\pm\omega_{i'}) \label{P0jfinal}\\
&\times\bigg|\overline{u}_{\qb's'}\,\vec{\gamma}\cdot\sum_{\qb''}\Big[
\vec{\mathcal{E}}_{i,\qb'-\qb''}
\,G_F(\qb'',\varepsilon_{q'}-\omega_i)\,\vec{\mathcal{E}}^{\pm}_{i',\qb+\qb''}
+\vec{\mathcal{E}}^{\pm}_{i',\qb'-\qb''}\,
G_F(\qb'',\varepsilon_{q'}\pm\omega_{i'})\,\vec{\mathcal{E}}_{i,\qb+\qb''}
\Big]\cdot\vec{\gamma}\,v_{\qb s}\bigg|^2, \nonumber
\end{align}
where we have defined $\vec{\mathcal{E}}^+_{i',\kb}\equiv\vec{\mathcal{E}}^*_{i',-\kb}$ and $\vec{\mathcal{E}}^-_{i',\kb}\equiv\vec{\mathcal{E}}_{i',\kb}$, while the prime in the summation symbol indicates that it is restricted to $\omega_{i'}<\omega_i$. Equation~(\ref{P0jfinal}) describes pair production (an electron of wave vector $\qb'$ and spin $s'$, combined with a positron of wave vector $\qb$ and spin $s$) for any arbitrary field comprising components of nondegenerate frequencies $\omega_i$.

\subsection{Alternative derivation of the pair-production rate in the interaction picture}
\label{general}

An alternative procedure to calculate the desired production rate consists in starting from the interaction Hamiltonian in the interaction picture $\hat{\mathcal{H}}^I_{\rm int}(t)=\ee^{\ii \hat{\mathcal{H}}_0 t/\hbar}\hat{\mathcal{H}}\ee^{-\ii \hat{\mathcal{H}}_0 t/\hbar}$. We are interested in obtaining the leading contribution to the probability amplitude connecting the initial fermionic vacuum state  $\ket{0}$ to a final pair state $\ket{p\qb s,e\qb's'}$, which we write as $C_{p\qb s,e\qb's'}=\bra{p\qb s,e\qb's'}\hat{\mathcal{S}}(\infty,-\infty)\ket{0}$ in terms of the scattering operator $\hat{\mathcal{S}}(t_2,t_1)=\mathcal{T}\,\ee^{-(\ii/\hbar)\int_{t_1}^{t_2}dt~\hat{\mathcal{H}}^I_{\rm int}(t)}$, where $\mathcal{T}$ stands for the time ordering operator. By retaining only quadratic terms in the electromagnetic field and working out time ordering through Wick's theorem \cite{MS10}, we obtain
\begin{align}
C_{p\qb s,e\qb's'}\approx&\frac{-\ii e^2}{\hbar^2}\int_{-\infty}^{\infty}\!\!\!\!\!\!dt\!\int_{-\infty}^\infty \!\!\!\!\!\!dt'\!\int \!\!d^3\rb \!\int \!\!d^3\rb'\, \bra{p\qb s,e\qb's'}:\overline{\Psi}(\rb,t)\vec{\gamma}\cdot\Ab(\rb,t) G_F(\rb-\rb',t-t')\vec{\gamma}\cdot\Ab(\rb',t')\hat{\Psi}(\rb',t'):\ket{0},\nonumber
\end{align}
where $G_F(\rb,t)=(2\pi)^{-4}\int_{-\infty}^{\infty} d\omega\int d^3\qb\,\ee^{\ii\qb\cdot \rb-\ii \omega t}G_F(\qb,\omega)$ is the real-spacetime representation of the Feynman propagator defined in Eq.~(\ref{GFF}). Plugging the vector potential defined in Eq.~(\ref{Art}) and carrying out the required Dirac matrix algebra, this expression reduces to
\begin{align}
C^\pm_{p\qb s,e\qb's'}\approx &\frac{2\pi\ii e^2c^2}{V^2 \hbar^2}
\sum_{ii'}{}^{\prime} \frac{1}{\omega_i\omega_{i'}}
\;\delta(\varepsilon_q+\varepsilon_{q'}-\omega_i\pm\omega_{i'}) \nonumber\\
&\times\overline{u}_{\qb's'}\,\vec{\gamma}\cdot\sum_{\qb''}\Big[
\vec{\mathcal{E}}_{i,\qb'-\qb''}
\,G_F(\qb'',\varepsilon_{q'}-\omega_i)\,\vec{\mathcal{E}}^{\pm}_{i',\qb+\qb''}
+\vec{\mathcal{E}}^{\pm}_{i',\qb'-\qb''}\,
G_F(\qb'',\varepsilon_{q'}\pm\omega_{i'})\,\vec{\mathcal{E}}_{i,\qb+\qb''}
\Big]\cdot\vec{\gamma}\,v_{\qb s},
\nonumber 
\end{align}
where $\vec{\mathcal{E}}_{i,\kb}$ is defined in Eq.~(\ref{Eik}) and the $\pm$ sign refers to channels involving either two frequencies of opposite sign ($+$) or two positive frequencies ($-$). Again, the prime in the summation symbol restricts it to $\omega_{i'}<\omega_i$ terms. Finally, the transition rate is obtained as $\Gamma_{p\qb s,e\qb's'}^{(2)}=|C^\pm_{p\qb s,e\qb's'}|^2/T$, where $T$ is the interaction time. This expression produces a squared $\delta$-function that we need to reinterpret by retaining one of such functions coming from one of the two $C^\pm_{p\qb s,e\qb's'}$ factors and then undoing the time integral in the other factor through the prescription $\delta\to(2\pi)^{-1}\int dt$; the remaining $\delta$-function still imposes energy conservation, whereas the undone time integral yields a factor $T$ that cancels with the denominator. Following this procedure, we readily find a result that coincides with Eq.~(\ref{P0jfinal}).

\section{Pair production by $\gamma$-ray interaction with a general polaritonic field}
\label{appendixB}
\renewcommand{\theequation}{B\arabic{equation}} 

Equation~(\ref{P0jfinal}) can be generally applied to an arbitrary number of field components. Here, we are interested in calculating the pair-production rate associated with the scattering of surface polaritons of frequency $\op$ ($i=p$) and highly energetic ($>2\me c^2\approx1.02\,$MeV) $\gamma$-ray photons of frequency $\og$ ($i=\gamma$). We consider a general polaritonic field $\vec{\mathcal{E}}_{p}(\rb)$, from which the Fourier-transformed field $\vec{\mathcal{E}}_{p,\kb}$ is obtained by using Eq.~(\ref{Eik}). Likewise, we write $\vec{\mathcal{E}}_\gamma(\rb)=E_\gamma\,\eh_j\,\ee^{\ii \kbg \cdot \rb }$ for a $\gamma$-ray plane-wave field of amplitude $E_{\gamma}$, wave vector $\kbg$ (taking along $\zz$), and unit polarization vector $\eh_j=\xx$ or $\yy$ for $j=1$ or 2, respectively, leading to $\vec{\mathcal{E}}_{\gamma\kb}=V\,E_\gamma\,\eh_j\,\delta_{\kb,\kbg}$. We neglect material polarization at the high $\gamma$-photon frequency. Inserting these expressions for the momentum-space fields into Eq.~(\ref{P0jfinal}), and noticing that the only term in the $(i,i')$ sum satisfying $\omega_{i'}<\omega_i$ corresponds to the choice $i=\gamma$ and $i'=p$, we find the rate
\begin{align}
\Gamma_{p\qb s,e\qb's'}^{(2)}=&\frac{\pi\alpha^2c^6|E_\gamma|^2}{V^2 \hbar^2\omega_\gamma^2\omega_p^2}\sum_\pm\delta(\varepsilon_q+\varepsilon_{q'}-\omega_\gamma\pm\omega_p)
\;\sum_{j=1,2}
\bigg|\overline{u}_{\qb's'}\,\mathcal{M}^\pm_j(\qb,\qb')\,v_{\qb s}\bigg|^2, \nonumber
\end{align}
where $\alpha\approx1/137$ is the fine-structure constant, the average over $\gamma$-ray polarization is performed ($j$ sum), we recall that primed (unprimed) quantities refer to the electron (positron), and we define the $4\times4$ matrix
\begin{align}
\mathcal{M}^\pm_j(\qb,\qb')=\gamma^j
\,G_F(\qb'-\kbg,\varepsilon_{q'}-\omega_\gamma)\,\vec{\gamma}\cdot\vec{\mathcal{E}}^{\pm}_{p,\qb+\qb'-\kbg}
+\vec{\mathcal{E}}^{\pm}_{p,\qb+\qb'-\kbg}\cdot\vec{\gamma}\,
G_F(\kbg-\qb,\omega_\gamma-\varepsilon_q)\,\gamma^j
\label{Mqq}
\end{align}
with $G_F$ given by Eq.~(\ref{GFF}).

It is convenient to recast this result in the form of a polariton-driven pair-production cross section $\sigma^{\rm pol}_{p\qb s,e\qb's'}=\Gamma_{p\qb s,e\qb's'}^{(2)}\big{/}N_pF_\gamma$, which is calculated by normalizing the rate to both the number of polaritons in the material ($N_p$) and the $\gamma$-photon flux traversing the polariton-supporting interface ($F_\gamma$). More precisely, we obtain $N_p$ as the space integral of the field energy density divided by the polariton energy,
\begin{align}
N_p=\frac{1}{4\pi\hbar\op}\int d^3 \rb \left\{\partial_{\omega_p}{\rm Re}\big\{\omega_p\epsilon(\rb,\omega_p)\big\}\,|\vec{\mathcal{E}}_p(\rb)|^2+(c/\op)^2|\nabla \times \vec{\mathcal{E}}_p(\rb)|^2\right\},
\label{Np}
\end{align}
where $\epsilon(\rb,\omega_p)$ is the position-dependent permittivity of the involved materials at the polariton frequency $\omega_p$. Here, the local response approximation is adopted and polaritons are assumed to be lossless as a reasonable description of long-lived modes. In addition, the $\gamma$-photon flux is derived from the associated intensity divided by the photon energy as $F_\gamma=c\,|E_\gamma|^2/2\pi\hbar\og$. Putting these elements together, we find the momentum-resolved positron-emission cross section
\begin{align}
\frac{d\sigma^{\rm pol}}{d\qb}=&\frac{V}{(2\pi)^3}\sum_{ss'}\sum_{\qb'}\sigma^{\rm pol}_{p\qb s,e\qb's'} \nonumber\\
=&\frac{\alpha^2c^5}{32\pi^4N_p\hbar\omega_\gamma\omega_p^2}
\int d^3\qb'
\sum_\pm\delta(\varepsilon_q+\varepsilon_{q'}-\omega_\gamma\pm\omega_p)
\sum_{ss'}\;\sum_{j=1,2}
\bigg|\overline{u}_{\qb's'}\,\mathcal{M}^\pm_j(\qb,\qb')\,v_{\qb s}\bigg|^2, \label{sgen}
\end{align}
which includes a sum over emitted-particle spins and incorporates the prescription $\sum_{\qb}\to(2\pi)^{-3}V\int d^3\qb$ to transform sums over particle wave vectors into integrals. The cross section in Eq.~(\ref{sgen}) is normalized in such a way that the total positron-emission cross section is given by $\sigma^{\rm pol}=\int d^3\qb\,(d\sigma^{\rm pol}/d\qb)$. Finally, the energy- and polar-angle-resolved positron-emission cross section is obtained by integrating Eq.~(\ref{sgen}) over the azimuthal emission angle $\varphi$ as
\begin{align}
\frac{d\sigma^{\rm pol}}{dE_qd\theta}=\sin\theta\,\frac{q\varepsilon_{q}}{\hbar c^2}\int_0^{2\pi} \!\!\!d\varphi\;\frac{d\sigma^{\rm pol}}{d\qb}, \label{dsdEdt}
\end{align}
where $E_q=\hbar\varepsilon_{q}$ is the positron energy and $\theta$ is the emission angle relative to the $z$ axis.

\section{Pair production by $\gamma$-ray interaction with a polaritonic gap mode}
\label{appendixC}
\renewcommand{\theequation}{C\arabic{equation}} 

Equation~(\ref{sgen}) gives the positron emission cross section for a general polaritonic field. Polaritonic gap modes are particularly interesting because they can enhance the optical field by several orders of magnitude relative to the incident light field within a small spatial region \cite{paper156}. To estimate the effect of field confinement and enhancement on the pair-production yield, we consider a mode field of uniform amplitude $E_p$ and unit polarization vector $\xx$ defined within a sphere of radius $R_p$, such that $\vec{\mathcal{E}}_p(\rb)=E_p\xx\,\Theta(R_p-r)$. The Fourier transform of this field (Eq.~(\ref{Eik})) yields
\begin{align}
\vec{\mathcal{E}}_{p,\kb}=\frac{4 \pi E_p \, \xx}{k^3}\left[\sin(kR_p)-kR_p\cos(kR_p)\right].\label{fgapk}
\end{align}
To compute the cross section per polariton and $\gamma$-photon, we plug Eq.~(\ref{fgapk}) into Eq.~(\ref{sgen}) and divide the result by the number of polaritons, which is obtained from Eq.~(\ref{Np}) as $N_p\approx E_p^2R_p^3/3\hbar \omega_p$ after disregarding the energy contribution from the magnetic field (this part is negligible for highly confined modes supported by the electric polarization of the involved materials). Finally, transforming the $\delta$ function in Eq.~(\ref{sgen}) as $\delta(\varepsilon_q+\varepsilon_{q'}-\omega_\gamma\pm\omega_p)=\delta(q'-q'_\pm)/\partial_{q'}\varepsilon_{q'}$ with $\partial_{q'}\varepsilon_{q'}=c^2q'/\varepsilon_{q'}$, we obtain
\begin{align}
\frac{d\sigma^{\rm pol}}{d\qb}=&\frac{3\alpha^2c^3}{2\pi^2\omega_\gamma\omega_p R_p^3}
\int d\Omega_{{\qb}'} 
\sum_\pm \left[\sin(k_{p\pm} R_p)-k_{p\pm} R_p\cos(k_{p\pm} R_p)\right]^2\Theta(\omega_\gamma \mp\omega_p-\me c^2/\hbar-\varepsilon_q)
 \label{sggap}\\
&\times\frac{q'_\pm\varepsilon_{q'_\pm}}{k_{p\pm}^6}\!\!
\sum_{ss'}\!\sum_{j=1,2} \bigg|\overline{u}_{\tilde{\qb}'_\pm s'}
\Big[\gamma^j\,
G_F({\qb}'_\pm-\kbg,\varepsilon_{q'_\pm}-\omega_\gamma)\,{\gamma}^1
+{\gamma}^1\,G_F(\kbg-\qb, \omega_\gamma-\varepsilon_q)\,\gamma^j
\Big]v_{\qb s}\bigg|^2, \nonumber
\end{align}
where $k_{p\pm}=|\qb+\qb'_\pm-\kb_\gamma|$, while $q'_\pm$ must satisfy the energy conservation condition $\varepsilon_{q'_\pm}=\omega_\gamma-\varepsilon_q\mp\omega_p$.

\setcounter{figure}{1}

\begin{figure*}[h]
\begin{centering} \includegraphics[width=0.8\textwidth]{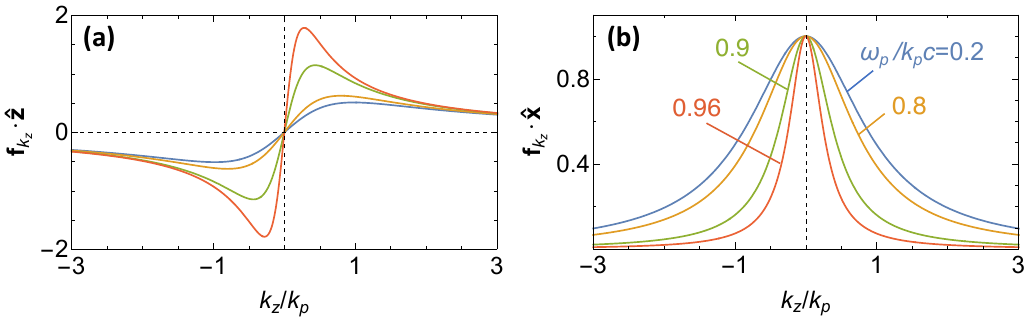} \par\end{centering}
\caption{{\bf Momentum distribution associated with the polariton field.} We plot the components of $\fb_{k_z}$ (Eq.~(\ref{fkz})) along (a) out-of-plane and (b) in-plane directions as a function of $k_z$ (normalized to the polariton wave vector $k_p$) for various polariton frequencies $\omega_p$ (normalized to $k_pc$).}
\label{FigS1}
\end{figure*}

\section{Pair production by $\gamma$-ray interaction with a surface polariton}
\label{appendixD}
\renewcommand{\theequation}{D\arabic{equation}} 

We consider polaritons bound to a planar material interface of area $A$ placed in the $z=0$ plane (e.g., a two-dimensional material capable of supporting strongly confined polaritons \cite{paper283,LCC17}, such as graphene \cite{paper235}, few-atomic-layer hexagonal boron nitride \cite{CAC19}, or ultrathin metal films \cite{paper335}). Polaritons are taken to be lossless and traveling with a real in-plane wave vector $\kb_p=\kp\xx$ (with $k_p>\omega_p/c$) oriented along the $x$ direction, so that their associated electric field can be written as $\vec{\mathcal{E}}_p(\rb)=(E_p c/\op) \big(\ii\kappa_p\,\xx -\kp\,{\rm sign}\{z\}\,\zz \big)\ee^{\ii \kp x -\kappa_p|z|}$, where $E_p$ is a global amplitude and $\kappa_p=\sqrt{k_p^2-\omega_p^2/c^2}$ describes the evanescent field decay away from the interface. The Fourier transform of this field (see Eq.~(\ref{Eik})) is
\begin{align}
\vec{\mathcal{E}}_{p\kb}&=\frac{2\ii A\,c}{\op}\,E_p\, \fb_{k_z} \,\delta_{\kb_\parallel,\kb_p}=\frac{2\ii\,c}{\op}\,E_p\, \fb_{k_z} \,(2\pi)^2\delta(\kb_\parallel-\kb_p), \label{Eikpg}
\end{align}
where the subscript $\parallel$ denotes the in-plane $x$-$y$ components, we define the real vector
\begin{align}
\fb_{k_z}&=\frac{\kappa_p^2\,\xx+\kp k_z\,\zz}{\kappa_p^2+ k_z^2}, \label{fkz}
\end{align}
and the transformation from the central to the rightmost parts of Eq.~(\ref{Eikpg}) is carried out by using the relation $(2\pi)^2\delta(\kb_\parallel-\kb_p)=A\,\delta_{\kb_\parallel,\kb_p}$ between the Kroneker and Dirac $\delta$ functions. The lack of translational invariance along the out-of-plane direction introduces a finite range of momentum mismatch in that direction relative to the $q_z+q'_z=k_{\gamma z}$ condition, as described by the $k_z$ dependence of $\fb_{k_z}$, which we illustrate in Fig.~\ref{FigS1} (see also Eq.~(\ref{Nqq}) below). From this polaritonic field, Eq.~(\ref{Np}) yields a number of polaritons $N_p=A\,|E_p|^2 k_p^2c^2/2\pi\hbar\omega_p^3\kappa_p$. Inserting this result together with Eq.~(\ref{Eikpg}) in Eq.~(\ref{sgen}), we find
\begin{align}
\frac{d\sigma^{\rm pol}}{d\qb}=\frac{\alpha^2c^5\kappa_p}{\pi\omega_p\omega_\gamma k_p^2}
\int d^3\qb'&\sum_\pm
\;\delta(\kb_{\gamma\parallel}-\qb_\parallel-\qb'_\parallel\mp\kb_p)
\;\delta(\varepsilon_q+\varepsilon_{q'}-\omega_\gamma\pm\omega_p) \label{sigmapppre}\\
&\times\sum_{ss'}\sum_{j=1,2}
\Big|\overline{u}_{\qb's'}\,\mathcal{N}^\pm_j(\qb,\qb')\,v_{\qb s}\Big|^2, \nonumber
\end{align}
where the $4\times4$ matrix
\begin{align}
\mathcal{N}^\pm_j(\qb,\qb')=\gamma^j
\,G_F(\qb'-\kb_\gamma,\varepsilon_{q'}-\omega_\gamma)\,\vec{\gamma}\cdot\fb_{\pm(k_{\gamma z}-q_z-q'_z)}
+\fb_{\pm(k_{\gamma z}-q_z-q'_z)}\cdot\vec{\gamma}\,
G_F(\kb_\gamma-\qb,\varepsilon_{q'}\pm\omega_p)\,\gamma^j
\label{Nqq}
\end{align}
is obtained from Eq.~(\ref{Mqq}) upon substitution of $\vec{\mathcal{E}}^{\pm}_{p,\qb+\qb'-\kbg}$ by $\fb_{\pm(k_{\gamma z}-q_z-q'_z)}$. The latter incorporates the anticipated finite out-of-plane momentum distribution. For a given emitted positron wave vector $\qb$, the electron wave vector $\qb'$ is determined by the $\delta$ functions in Eq.~(\ref{sigmapppre}). In particular, the in-plane electron wave vector is given by $\qb'_{\parallel\pm}=\kb_{\gamma\parallel}-\qb_\parallel\mp\kb_p$. Also, noticing the relation $\partial_{q'_z}\varepsilon_{q'}=q'_zc^2/\varepsilon_{q'}$, we can write
\begin{align}
\delta(\varepsilon_q+\varepsilon_{q'}-\omega_\gamma\pm\omega_p)
=\frac{\varepsilon_{q'_\pm}}{c^2q'_{z\pm}}\;
\big[\delta(q'_z-q'_{z\pm})+\delta(q'_z+q'_{z\pm})\big]
\;\Theta\Big(\varepsilon_{q'_\pm}^2-\me^2c^4/\hbar^2-c^2q_{\parallel\pm}^{\prime\,2}\Big)
\;\Theta\Big(\og\mp\op-\varepsilon_{q}\Big),
\nonumber
\end{align}
where
\begin{align}
q'_{z\pm}=\sqrt{\varepsilon_{q'_\pm}^2/c^2-\me^2c^2/\hbar^2-q_{\parallel\pm}^{\prime\,2}}
\label{qparallel}
\end{align}
is the out-of-plane electron wave-vector component and $\varepsilon_{q'_\pm}=\og\mp\op-\varepsilon_{q}$ is the electron energy. This allows us to recast the wave-vector-resolved differential positron emission cross section as
\begin{align}
\frac{d\sigma^{\rm pol}}{d\qb}=&\frac{\alpha^2c^3\kappa_p}{\pi\,\omega_p\omega_\gamma k_p^2}
\sum_\pm
\frac{\varepsilon_{q'_\pm}}{q'_{z\pm}}
\;\Theta\Big(\varepsilon_{q'_\pm}^2-\me^2c^4/\hbar^2-c^2q_{\parallel\pm}^{\prime\,2}\Big)
\;\Theta\Big(\og\mp\op-\varepsilon_{q}\Big) \label{sigmapp}\\
&\quad\quad\quad\quad\quad\quad\quad \times
\sum_{ss'}\;\sum_{j=1,2} \; \sum_{\mu=\pm1} \;
\Big|\overline{u}_{\qb'_{\parallel\pm}+\mu q'_{z\pm}\zz,s'}\,\mathcal{N}^\pm_j(\qb,\qb'_{\parallel\pm}+\mu q'_{z\pm}\zz)\,v_{\qb s}\Big|^2, \nonumber
\end{align}
with $q'_{z\pm}$ given in Eq.~(\ref{qparallel}), such that the $\mu=1$ and $\mu=-1$ terms stand for the contributions associated with upward ($q'_z=+q'_{z\pm}$) and downward ($q'_z=-q'_{z\pm}$) electron emission, respectively.

Finally, we insert Eq.~(\ref{sigmapp}) into Eq.~(\ref{dsdEdt}) to compute $d\sigma^{\rm pol}/dE_qd\theta$ in Figs.~\ref{FigS2} and \ref{FigS3} below, where we present this quantity after averaging it over a finite positron energy range $\Delta E_q=1$~keV just to make the plot clearer by smoothing the integrable divergence introduced by the $1/q'_{z\pm}$ factor in Eq.~(\ref{sigmapp}) at the onset of positron emission.

\begin{figure*}
\begin{centering} \includegraphics[width=1.0\textwidth]{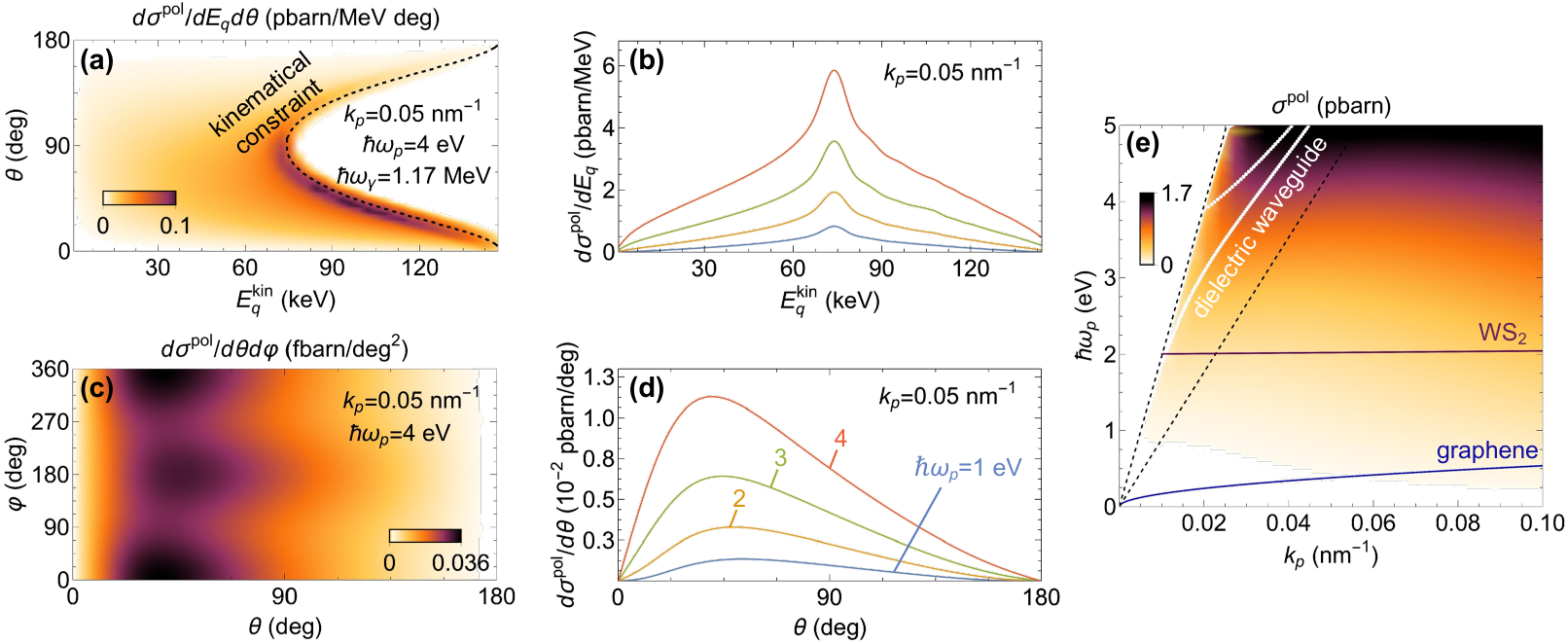} \par\end{centering}
\caption{{\bf Pair-production cross section near the threshold.} (a)~Differential cross section for positron emission as a function of polar angle $\theta$ and kinetic energy $E^{\rm kin}_{q}=\hbar \varepsilon_{q}-\me c^2$ (normalized to the $\gamma$-photon energy $\hbar \og=1.17$~MeV and averaged over a window $\Delta E^{\rm kin}_q=8$~keV) for fixed polariton wave vector $\kp=0.05$~nm$^{-1}$ and energy $\hbar\omega_p=4$~eV, as computed from $d\sigma^{\rm pol}/dE_qd\theta=\sin\theta\,(q\varepsilon_{q}/\hbar c^2)\int_0^{2\pi} \!\!\!d\varphi\;(d\sigma^{\rm pol}/d\qb)$ with the integrand taken from Eq.~(\ref{sigmapp}). The dashed line represents the limit imposed by energy-momentum conservation for $\varphi=0$. (b)~Same as (a) integrated over $\theta$ for different polariton energies $\hbar\omega_p$ (see color labels in (d)). (c)~Energy-integrated cross section $d\sigma^{\rm pol}/d\theta d\varphi=\sin\theta\,\int_0^\infty q^2dq\,(d\sigma^{\rm pol}/d\qb)$ as a function of polar and azimuthal emission angles $(\theta,\varphi)$ under the conditions of (a). (d)~Same as (c) integrated over $\varphi$ for different polariton energies. (e) Total cross section ($\qb$-integral of Eq.~(\ref{sigmapp})) as a function of polariton wave vector $\kp$ and energy $\hbar\op$.
For reference, we show the dispersion relations of free-space light ($\omega=ck$), p-polarized modes in a dielectric waveguide \cite{notepositron5} (80~nm thickness, 2.24 refractive index), graphene plasmons (1\,eV Fermi energy), and the A exciton in monolayer WS$_2$ \cite{notepositron4}.}
\label{FigS2}
\end{figure*}

\subsection{Pair production close to the threshold}

From Eq.~(\ref{sigmapp}) and the discussion presented in the main text, we expect positron production by mixing polaritons and $\gtrsim1.02$~MeV photons, such as those available from commonly used sources \cite{APP14,DFH14} (e.g., $^{60}$Co, which emits at $\sim1.17$~MeV and $\sim1.33$~MeV with a lifetime of $\sim5.13$ years, yielding $\sim10^{14}$ photons/s out of 1~g of material).

To put the present results in context, we note that the free-space Breit-Wheeler (BW) cross section \cite{notepositron2} is very small for pair production out of such $\gamma$-photons alone (e.g., the maximum cross section is $\sigma^{\rm BW}\lesssim0.17$~barn ($1$ barn $= 10^{-24}$ cm$^2$) for two 1.33~MeV photons). We illustrate this by considering an arrangement consisting of two facing ${}^{60}$Co sources spaced by a few meters so that $\sim10^6$ photons are simultaneously traveling across that distance, and therefore, $\sim10^{12}$ photon-photon collisions take place during the traveling time $\sim10^{-8}$~s. Now, multiplying the number of collisions by $\sigma^{\rm BW}$ and dividing by both a transverse area of $\sim1$~m$^2$ and the traveling time, we estimate a pair-production rate of $\sim10^{-9}$/s.

Polaritons can be made in large supply over small spatial regions by relying on ultrafast lasers (e.g., one has $\sim10^{19}$ photons in 1~J pulses of 100~fs duration, such as those delivered by tabletop setups, which could be schemed to achieve nearly complete coupling to polaritons \cite{paper370}). This allows us to partially compensate for the small polariton-induced pair-production cross section at such relatively small $\gamma$-photon energies (e.g., $\sigma^{\rm pol}\sim 0.1$~pbarn for few-eV polaritons and 1.17~MeV $\gamma$-photons; see Fig.~\ref{Fig2}c in the main text). For example, considering again $\gamma$-photons delivered by a ${}^{60}$Co source close to a polariton-supporting surface, we can have a flux of $10^{14}$~$\gamma$-photons/s~cm$^2$, which, when multiplied by $\sigma^{\rm pol}$, by the number of polaritons $N_p\sim10^{19}$, and by the polariton lifetime (e.g., nanoseconds for high-index planar dielectric waveguides with quality factors $\sim10^6$), leads to $\sim10^{-13}$ pairs per laser pulse, which can be collected over a time measurement window of $\sim0.1~$ns using currently available fast electronics.

Considering the use of these kinds of sources, we consider $\hbar \og=1.17$~MeV and compute the emitted positron distribution predicted by Eq.~(\ref{sigmapp}) as a function of kinetic energy $E^{\rm kin}_{q}=\hbar\varepsilon_{q}-\me c^2$ and polar angle $\theta$ under the configuration depicted in Fig.~\ref{Fig2}a of the main text. The result (Fig.~\ref{FigS2}a) indicates a preference for polar angles close to normal when the positron takes most of the energy (electron emitted nearly at rest), and conversely, grazing emission for low-energy positrons. The spectral distribution obtained by further integrating over $\theta$ displays a symmetric profile with respect to the central peak found at $E^{\rm kin}_{q}=(\hbar\og-2\me c^2)/2\approx74$~keV (Fig.~\ref{FigS2}b), as expected from the electron-positron kinematical symmetry. In addition, the energy-integrated positron-emission cross section is nearly independent of azimuthal angle $\varphi$ (Fig.~\ref{FigS2}c) because of the comparatively small in-plane momentum carried by the polaritons, while the polar dependence shows a maximum at around $\theta\sim 45^\circ$, in good correspondence with the symmetrically arranged pair emission, dominated by the spectral maximum in Fig.~\ref{FigS2}b. Finally, the full $\qb$-integrated cross section (Fig.~\ref{FigS2}e) shows a nearly uniform increase with polariton frequency as $\propto\omega_p^2$, except for the depletion observed when $k_p$ moves close to the light cone (dashed line). Overall, we conclude that the studied process leads to a strong angular and energy dependence of the resulting positron emission, which should facilitate an experimental verification of these results, although the background coming from Bethe-Heitler scattering at the polaritonic material imposes a severe constrain, as we discuss in the main text.

\begin{figure}
\begin{centering} \includegraphics[width=0.85\textwidth]{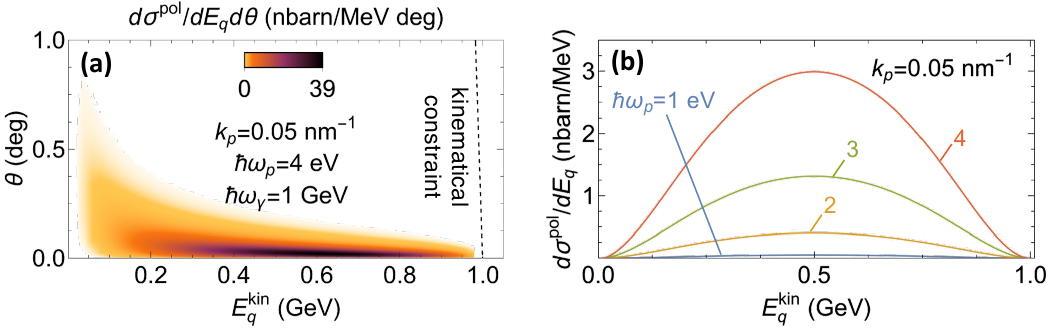} \par\end{centering}
\caption{{\bf Pair production with polaritons and GeV $\gamma$-photons.} (a) Pair-production differential cross section as a function of positron polar angle $\theta$ and kinetic energy $E^{\rm kin}_{q}=\hbar \varepsilon_{q}-\me c^2$ (normalized to the $\gamma$-photon energy $\hbar \og=1$~GeV and averaged over an energy window $\Delta E^{\rm kin}_q=1$~keV) for fixed polariton momentum $\kp=0.05$~nm$^{-1}$ and energy $\hbar\op=4$~eV. The dashed line represents the limit imposed by energy-momentum conservation for an azimuthal angle $\varphi=0$. (b)~Spectral distribution of positron emission (integrated over $\theta$) for different polariton energies.}
\label{FigS3}
\end{figure}

\subsection{Pair production with $\text{GeV}$ $\gamma$-photons}

We analyze the emitted positron distribution in Fig.~\ref{FigS3} for 1~GeV $\gamma$-photons, which can be experimentally produced by bremsstrahlung and Compton backscattering \cite{NTS04}, while several proposals for more efficient sources have recently been put forward based on electron-beam collisions with intense laser spots \cite{GBB17,MGM19}, strong laser irradiation of electron plasma \cite{LSZ18,ZYC18}, simultaneous laser and electron plasma bombardment \cite{ZCW20}, and electrons impinging on solid targets \cite{SDC21}.

Upon integration of Eq.~(\ref{sigmapp}) over the azimuthal positron emission angle $\varphi$, Fig.~\ref{FigS3}a illustrates how the differential cross section $d\sigma^{\rm pol}/dE_{q}d\theta$ (see Eq.~(\ref{dsdEdt})) is strongly peaked around normal emission (polar angle $\theta\sim0$). In addition, positrons are preferentially sharing about half of the photon energy (Fig.~\ref{FigS3}b), with a similar spectral distribution regardless of the polariton energy and a strong increase in emission efficiency with polariton energy $\hbar\omega_p$ (already observed in Fig.~\ref{FigS2}b for $1.17$~MeV $\gamma$-photons).

We remark that polaritonic modes can be strongly populated by irradiation with ultrafast laser pulses at fluences creating a surface polariton density as high as $\rho_p\sim1/$nm$^2$ without causing material damage, such that the scattering of 1~GeV photons ($\sigma^{\rm pol}\sim10^{-6}$~barn; see Fig.~\ref{Fig2}c in the main text) at a currently attainable rate $r_\gamma\sim10^6$/s \cite{MYY22} would lead to a pair-production rate $\rho_pr_\gamma\sigma^{\rm pol}\sim10^{-10}$/s, while higher rates could potentially be achieved with alternative designs for efficient GeV photon sources \cite{GBB17,ZCW20}.

\end{widetext}

\section*{References}
\renewcommand{\bibsection}{}


\begin{thebibliography}{61}
\bibitem [{not({\natexlab{e}})}]{notepositron5}%
  \BibitemOpen
  \href@noop {} {} \bibinfo {note} {The dispersion relation
  of p-polarized guided modes in a self-standing dielectric waveguide with real
  refractive index $n=\sqrt{\epsilon}$ and thickness $d$ is computed
  numerically by solving the equations $\cot(k_z d/2)=-\epsilon \kappa_z/k_z$
  and $\tan(k_z d/2)=\epsilon \kappa_z/k_z$, with
  $k_z=\sqrt{\omega_p^2/c^2-k_p^2}$ and $\kappa_z=\sqrt{k_p^2-\omega_p^2
  \epsilon/c^2}$ \cite{paper370}.}\BibitemShut {Stop}%
\bibitem [{not({\natexlab{f}})}]{notepositron4}%
  \BibitemOpen
  \href@noop {} {} \bibinfo {note} {The quasistatic
  dispersion relation of a surface polariton propagating along a 2D material
  with conductivity $\sigma(\omega)= ( \ii
  e^2/\hbar)\,\omega_D/(\omega-\omega_g+\ii\gamma)$ is
  $\omega_p=\omega_g/2+\sqrt{\omega_g^2/4+2\pi e^2 \omega_D k_p\hbar}$, as
  obtained from $\omega_p=-2\pi\ii\sigma k_p$ \cite{paper235}. For graphene
  plasmons, we set $\omega_g=0$ and $\hbar \omega_D=\EF/\pi$, whereas for
  WS$_2$ excitons, we set $\hbar\omega_g=2$ eV and $\hbar\omega_D= 100$ meV
  \cite{LCZ14}.}\BibitemShut {Stop}%
\end{thebibliography}

%

\section*{ACKNOWLEDGMENTS}

This work has been supported in part by the European Research Council (Advanced Grant 789104-eNANO), the European Commission (Horizon 2020 Grants FET-Proactive 101017720-EBEAM and FET-Open 964591-SMART-electron), the Spanish MICINN (PID2020-112625GB-I00 and Severo Ochoa CEX2019-000910-S), the Catalan CERCA Program, and Fundaci\'os Cellex and Mir-Puig.

\end{document}